\documentclass[prb,showpacs,amssymb]{revtex4}

\usepackage{graphicx}

\begin{document}

\title{Superconductors with broken time-reversal symmetry:
Spontaneous magnetization and quantum Hall effects
\\}

\author{Baruch Horovitz$^{1,2}$ and Anatoly Golub$^1$}
\affiliation{$^1$Department of Physics and $^2$Ilze Katz center for
nanotechnology, Ben-Gurion University of the Negev, Beer-Sheva
84105, Israel}
\begin{abstract}

Broken time reversal symmetry (BTRS) in $d+id'$ as well as in
$d+is$ superconductors is studied and is shown to yield current
carrying surface states. We evaluate the temperature and thickness
dependence of the resulting spontaneous magnetization and show a
marked difference between weak and strong BTRS. We also derive the
Hall conductance which vanishes at zero wavevector $q$ and finite
frequency $\omega$, however at finite $q, \omega$ it has an
unusual structure. The chirality of the surface states leads to
quantum Hall effects for spin and heat transport in $d+id'$
superconductors.
\end{abstract}


\maketitle


\section{Introduction}

Recent data on the high $T_c$ superconductor $YBa_2Cu_3O_x$ (YBCO)
has supported the presence of broken time reversal symmetry (BTRS)
\cite{covington,deutscher,polturak}.  A sensitive probe of BTRS
are Andreev surface states. For a d wave with time reversal
symmetry bound states at zero energy are expected for a surface
parallel to the nodes (i.e. a (110) surface in YBCO).  When BTRS
is present, by either a complex order parameter or by an external
magnetic field, the bound states shift to a finite energy. Indeed
tunnelling data usually shows a zero bias peak which splits in an
applied field; the splitting is nonlinear in the magnetic field,
indicating a proximity to a BTRS state \cite{deutscher,alff}. In
fact, in some samples tunnelling data shows a splitting even
without an external field \cite{covington,deutscher}, consistent
with BTRS; the splitting increases with increasing overdoping
\cite{deutscher,sharoni},

Further support for a spontaneous BTRS state are spontaneous
magnetization data as observed in YBCO \cite{polturak}, setting in
abruptly at $T_c$ and being almost temperature ($T$) independent
below $T_c$.  The phenomenon has been attributed to either a
$d_{x^2-y^2}+id_{xy}$ state ($d+id'$) or to formation of $\pi$ junctions.
No microscopic reason was given, however, for the spontaneous
magnetization being independent of both $T$ and of
 film thickness \cite{polturak}.

It has been shown theoretically that BTRS can occur locally in a
$d_{x^2-y^2}$ superconductor near certain surfaces
\cite{sigrist1,matsumoto,rainer,kos} leading to either $d+id'$ or
$d+is$ states with surface currents.  The
onset of such BTRS is expected to be below $T_c$ and therefore
does not correspond to the spontaneous magnetization data
\cite{polturak}.  We note that in response to an external magnetic
field the surface states are paramagnetic and compete with
Meissner currents. This effect has been proposed to account for a
minimum in the magnetic penetration length \cite{walter}. In fact,
it was proposed that this paramagnetic effect leads to
spontaneous currents and BTRS in a pure $d_{x^2-y^2}$ state
\cite{higashitani,barash}. The onset of this BTRS is much below
$T_c$ \cite{barash} and therefore does not correspond to the data
\cite{polturak}.

Of further theoretical interest is the relation of the BTRS state
to quantum Hall systems with a variety of Hall effects
\cite{goryo,horovitz,senthil,read,goryo2}.  In particular a finite charge
hall conductance has been suggested \cite{goryo}, though this has
been questioned \cite{read}.

In the present work we expand our earlier work \cite{golub} and
study variety of phenomena related to surface currents.
In section II we show that bulk $d+id'$ state has surface states with finite surface
currents; a similar situation was found for the bulk p wave state
\cite{sigrist2}. We also consider a $d+is$ state which has surface
currents only on the (110) surface. In section III we evaluate the
spontaneous magnetization and show that for $d+id'$ it is
dominated by $(100)$ surfaces; for thin films it increases with
the ratio
 $\Delta '/\Delta$ ($\Delta$ and $\Delta '$ are the amplitudes of
$d_{x^2-y^2}$ and $d_{xy}$, respectively) while for thick films it
has a maximum at $\lambda/\xi ' \approx 1$ where
$\xi'=v_F/\Delta'$ with $v_F$ the Fermi velocity
 and $\lambda$ is the penetration length, i.e.
at $\Delta '/\Delta \approx 0.01$ for YBCO. We show that for weak
BTRS, $\lambda/\xi'<1$, the spontaneous magnetization is $T$ and
thickness independent, while for strong BTRS thickness and $T$
dependence may occur.
 For the sample of Ref
\onlinecite{polturak} we estimate $\Delta '/\Delta \approx
10^{-4}$, i.e. weak BTRS. In section IV we consider a surface
approach for the quantum Hall effect, showing quantization for
spin and thermal Hall conductances for the $d+id'$ state. We also derive in section V
the effective action in the bulk and identify the Hall coefficient
which has an unusual wavevector and frequency dependence.

\section{Surface states}

We present here the Bogoliubov de-Gennes (BdG) equations for
quasiparticles in a bulk $d+id'$ or $d+is$ states
in presence of a boundary and study the resulting
surface states. We consider first a $d+id'$ state
where the order parameter is
\begin{equation}
 \Delta ({\hat p}_x,{\hat p}_y)= \Delta({\hat p}_x^2-{\hat p}_y^2)/k_F^2
 +i\Delta '{\hat p}_x{\hat p}_y/k_F^2
 \end{equation}
where ${\hat {\bf p}}=-i\hbar {\mbox{\boldmath $\nabla$}}$ is the
momentum operator and $k_F$ is the Fermi momentum. The
quasiparticles are represented by an electron-hole Nambu spinor
\begin{eqnarray}\label{spinor}
\Psi ({\bf r})=\left( \matrix{ \Psi_{\uparrow}({\bf r}) \cr \cr
\Psi^{\dagger}_{\downarrow}({\bf r})} \right)
\end{eqnarray}
and are described by the following mean field Hamiltonian (see
appendix A)
\begin{equation}
{\hat {\cal H}}=\frac{1}{2m}[(i{\mbox{\boldmath
$\nabla$}}+\frac{e}{c}\tau_3{\bf A}({\bf r}))^2-k_F^2]\tau_3+
\left(\matrix{ 0  &  \Delta ({\hat p}_x,{\hat p}_y) \cr \cr
\Delta^* ({\hat p}_x,{\hat p}_y) & 0} \right)e^{-i\theta ({\bf
r})\tau_3}
\end{equation}
where $m$ is the electron mass and $\tau_i$ are the Pauli
matrices. We assume here that $|{\mbox{\boldmath $\nabla$}}\theta|
\ll k_F$ so that the issue of gauge invariance in the interaction
term can be avoided (appendix A). Rotating by the unitary
transformation $\Psi({\bf r})\rightarrow exp[i\tau_{3} \theta
({\bf r})/2] \Psi({\bf r})$ yields
\begin{equation}
{\hat {\cal
H}}=\frac{1}{2m}(-\nabla^2-k_F^2)\tau_3+\frac{1}{2m}{\bf p}\cdot
({\mbox{\boldmath $\nabla$}}\theta -\frac{2e}{c}{\bf A}({\bf r}))
+ \left(\matrix{ 0  &  \Delta ({\hat p}_x,{\hat p}_y) \cr \cr
\Delta^* ({\hat p}_x,{\hat p}_y) & 0} \right)
\end{equation}
where ${\bf A}$ is kept to first order.

We consider a vacuum-superconductor boundary at x=0, and assume
for now that $\Delta, \Delta '$ are constants at $x>0$ and vanish
at $x<0$. For $\Delta > \Delta '$ this corresponds to a $(100)$
surface; to
 describe a $(110)$ surface $\Delta$ and $\Delta '$ need to be
 interchanged. The
spinor wavefunctions for the up and down component of Eq.
(\ref{spinor}), respectively, $u({\bf r})=u\exp [ifx+ik_yy]$ and
$v({\bf r})=v\exp [ifx+ik_yy]$ with eigenvalues $\epsilon$ satisfy
the BdG equations
\begin{eqnarray}
(f^2-k_F^2+k_y^2-2m{\tilde
\epsilon})u+2m\Delta(f,k_y)v=&&0\nonumber\\
(-f^2+k_F^2-k_y^2-2m{\tilde \epsilon})v+2m\Delta^*(f,k_y)u=&&0
\end{eqnarray}
where ${\tilde \epsilon}=\epsilon+(e/mc)k_yA_y(x)$, ${\bf A}$ has
only an $A_y$ component consistent with a current in the $y$
direction and ${\mbox{\boldmath $\nabla$}}\theta=0$. This Doppler
shift assumes that $A_y(x)$ is slowly varying on the scale
$k_F^{-1}$ so that a local eigenevalue ${\tilde \epsilon}$ can be
defined. Define $k=+\sqrt{k_F^2-k_y^2}$, then $f$ has two surface
solutions with $\Im f >0$
\begin{eqnarray}\label{f12}
f_1=&&k+i\frac{m}{k}\sqrt{|\Delta(k,k_y)|^2-{\tilde
\epsilon}^2}\nonumber\\
f_2=&&-k+i\frac{m}{k}\sqrt{|\Delta(-k,k_y)|^2-{\tilde \epsilon}^2}
\end{eqnarray}
where the replacement $\Delta(f,k_y)\rightarrow \Delta(\pm k,k_y)$
is valid for $|\Delta|,{\tilde \epsilon}\ll k_F^2/2m$. The
eigenvectors are
\begin{eqnarray}\label{v12}
v_1&=&-i \frac{\sqrt{|\Delta(k,k_y)|^2-{\tilde
\epsilon}^2}+i{\tilde \epsilon}}{\Delta(k,k_y)}u_1\nonumber\\
v_2=&&i\frac{\sqrt{|\Delta(-k,k_y)|^2-{\tilde
\epsilon}^2}-i{\tilde \epsilon}}{\Delta(-k,k_y)}u_2
\end{eqnarray}
We assume specular reflection which preserves $k_y$ but mixes
these two solutions so that at $x=0$ the wavefunctions vanish. A
linear combination for which both spinor components vanish at $x=0$, i.e.
$\alpha u_1+\beta u_2=\alpha v_1+\beta v_2=0$ yields
$v_1/u_1=v_2/u_2$, hence an equation for the eigenvalues
\begin{equation}\label{eigen0}
\frac{i{\tilde \epsilon} + \sqrt{|\Delta(+ k,k_{y})|^{2}-{\tilde
\epsilon}^{2}}} {-i{\tilde \epsilon} + \sqrt{|\Delta(-
k,k_{y})|^{2}-{\tilde \epsilon}^{2}}}= -\frac{\Delta(+
k,k_y)}{\Delta(- k,k_y)}
\end{equation}
Its solutions  are readily seen to be ${\tilde \epsilon} =-{\mbox
sign} (k_y)\Delta(k^2-k_y^2)/k_F^2$. In terms of the incidence
angle $\zeta$, $k_y=k_F\sin \zeta$, $k=k_F \cos \zeta$, the
eigenvalues are
\begin{equation}\label{eigen}
\epsilon_{\zeta}=-{\mbox sign} (\zeta)\Delta \cos
(2\zeta)-\frac{e}{c}v_FA_y \sin \zeta
\end{equation}
Note that the spectrum is not symmetric in $k_y$ or in $\zeta$ (it
is in fact antisymmetric) resulting in a finite surface current.
Fig. 1 shows the angle $\zeta$ where $\epsilon_{\zeta} =0$ (full lines)
and the range for which $\epsilon_{\zeta} >0$. The velocities $\partial
\epsilon_{\zeta} /\partial k_y$ are positive for both $\pm k_y$ branches,
i.e. the surface states are chiral. This property leads to
quantization of Hall effects, as discussed in section IV.

\begin{figure}
\begin{center}
\includegraphics[scale=0.5]{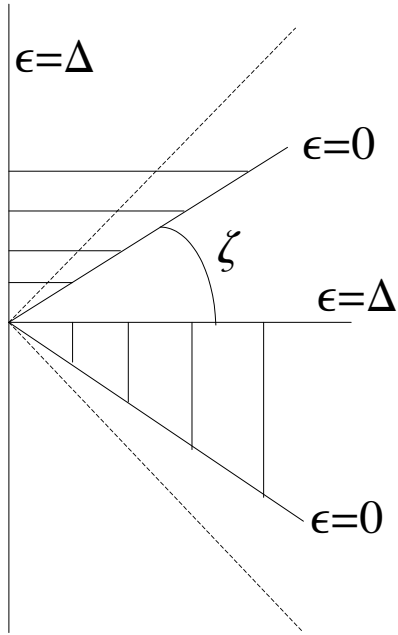}
\end{center}
\caption{Surface (vertical line) and angle $\zeta$ where the
spectrum is $\epsilon =0$ (full lines); in the absence of $A_y$
these lines would be at $\zeta =\pm \pi/4$ (dashed lines). The
range for which $\epsilon >0$ is shown as the hatched area; the
spectra spans the range $\epsilon =0$ up to $\epsilon = \Delta$ as
shown}
\end{figure}
\hspace{15mm}

We note that self consistency would imply that $\Delta '=0$ at
$x=0$ \cite{rainer}; the eigenfunctions would then be $\sim
\exp[-\int_0^x\Delta '(x')dx'|\sin \zeta|/v_F]$, resulting in a
very similar dependence on $\xi '$. Note also that
quasiparticles in the bulk have a spectral gap $|\Delta(k,k_y)|$ which
for any given $\zeta$ is higher than the surface states
$\epsilon_{\zeta}$ (neglecting the Doppler shift). Impurities,
however, may destroy $k_y$ conservation and scatter high energy
($>\Delta'$) surface states into degenerate bulk states. When impurity
scattering is essential (e.g. section IV) our results apply only
when these excitations can be neglected, e.g. at $T<\Delta'$. This
restriction is not needed at the $(110)$ surface where the whole
surface spectrum is $<\Delta'$, i.e. below the lowest bulk state.

The decay length of the surface states becomes, using Eqs.
(\ref{f12}, \ref{eigen}), $(\Im
f)^{-1}=[(m/k)\sqrt{|\Delta|^2-{\tilde
\epsilon}^2}]^{-1}=\xi'/|\sin \zeta|$ with $\xi'=v_F/\Delta'$.
Since $|u_i|=|v_i|, i=1,2$ (Eq. \ref{v12}) the normalized
eigenfunctions are
\begin{eqnarray}\label{unorm}
u_{\zeta}({\bf r})=&&\sqrt{\frac{2|\sin \zeta|}{\xi'L_y}}\sin kx
e^{ik_yy-x|\sin \zeta |/\xi'}\nonumber\\
v_{\zeta}({\bf r})=&&-{\mbox sign}(k_y)u({\bf r})
\end{eqnarray}
where $L_y$ is the length of the surface. It is remarkable that
$|u_{\zeta}({\bf r})|=|v_{\zeta}({\bf r})|$ for all $\zeta$, i.e.
for all energies of the surface states, implying maximal
electron-hole mixing. As noted above, a (110) surface has the same
solution (\ref{unorm}) with $\xi'$ replaced by $\xi=v_F/\Delta$.

We note that in general the spinor Eq. (\ref{spinor}) can be
decomposed in terms of eigenoperators $\eta_{\zeta \uparrow},
\eta_{\zeta \downarrow}$ where
\begin{eqnarray}\label{gammaspinor}
\left( \matrix{ \Psi_{\uparrow}({\bf r}) \cr \cr
\Psi^{\dagger}_{\downarrow}({\bf r})}
\right)=\sum_{\zeta}\left(\matrix{ u_{\zeta}({\bf r})  &
-v^*_{\zeta}({\bf r})\cr \cr v_{\zeta}({\bf r}) & u^*_{\zeta}({\bf
r})} \right)\left( \matrix{ \eta_{\zeta \uparrow} \cr \cr
\eta^{\dagger}_{\zeta \downarrow}} \right)
\end{eqnarray}
leading to the diagonal Hamiltonian
\begin{equation}\label{Hdiag}
{\hat {\cal H}}=\sum_{\zeta}\int dx
\epsilon_{\zeta}[\eta^{\dagger}_{\zeta \uparrow} \eta_{\zeta
\uparrow}+\eta^{\dagger}_{\zeta \downarrow} \eta_{\zeta
\downarrow}-1][|u(x)|^2+|v(x)|^2]
\end{equation}
with $\epsilon_{\zeta}$ being $x$ dependent via the Doppler shift.
The spectrum has exact particle-hole symmetry, i.e. for each
eigenvector $u,v$ with eigenvalue $\epsilon$ there is an
eigenvector $-v*,u*$ with eigenvector $-\epsilon$. The form Eq.
(\ref{Hdiag}) incorporates, however, both $\pm \epsilon$ states
and its sum is therefore restricted to $\epsilon_{\zeta}\geq 0$.

We consider next a $d+is$ state at a (110) surface with an order
parameter
\begin{equation}\label{dis}
\Delta ({\hat p}_x,{\hat p}_y)= \Delta {\hat p}_x{\hat
p}_y/k_F^2+i\Delta_s \,.
\end{equation}
Eq. (\ref{eigen0}) has then the solution ${\tilde \epsilon}={\mbox
sign}(k_y)\Delta_s$, i.e.
\begin{equation}\label{Ds}
\epsilon={\mbox sign}{\zeta}\Delta_s-\frac{e}{c}v_FA_y \sin
\zeta\,.
\end{equation}

Positive eigenvalues are now at $k_y\geq 0$ (for weak Doppler
effect $\frac{e}{c}v_F|A_y|<\Delta_s$) with a weak dispersion due
to the Doppler term. Note in particular that the spectrum has a gap,
i.e.  no
$\epsilon=0$ states; hence to probe these states one needs either high
voltage or high temperature $T>\Delta_s$. This $d+is$ state corresponds to a (110)
surface at which it breaks both parity and time reversal. At a
(100) surface the state $d+is$ state is symmetric under
reflection and in fact has no surface bound states. Hence
tunnelling data at the (100) may distinguish between $d+id'$ and
$d+is$ states, i.e. the $d+id'$ state shows a weak structure at a bias
$\approx \Delta$ while a $d+is$ state has no effect at all. The
magnetization data \cite{polturak} shows an effect for both (110) and
(100) surfaces, supporting a $d+id'$ state for YBCO.

\section{Spontaneous Magnetization}

The $d+id'$ or $d+is$ order parameters break both time reversal
invariance and reflection along the surface, hence they allows
surface currents ($d+is$ refers to (110) only). The current density parallel to a surface
(the $y$ direction) and the charge density are,
\begin{eqnarray}\label{edge}
j_{edge}(x)=&&\frac{-i\hbar e}{2md}\sum_s[\langle
\Psi_s^{\dagger}({\bf r})\frac{\partial}{\partial y}\Psi_s({\bf
r})\rangle-h.c.]=\frac{-2\hbar e}{md}\sum_{\zeta}k_y|u({\bf
r})|^2\tanh (\frac{\epsilon_{\zeta}}{2T})\nonumber\\
n_{edge}=&&\frac{e}{d}\sum_s\langle \Psi_s^{\dagger}({\bf
r})\Psi_s({\bf r})\rangle=\frac{2e}{d}\sum_{\zeta}|u_{\zeta}({\bf
r})|^2
\end{eqnarray}
where $d$ is the interlayer spacing, $\langle
\gamma_{\zeta,s}^{\dagger}\gamma_{\zeta,s}\rangle=[1+\exp (
\epsilon_{\zeta}/T)]^{-1}$ and $|u_{\zeta}({\bf r})|=|v_{\zeta}({\bf
r})|$ were used. The expression for $j_{edge}$ can also be
obtained from Eq. (\ref{Hdiag}) by $j_{edge}=c\delta H/\delta
A_y(x)$. In addition to the  explicit $T$ dependence in (\ref {edge})
the order parameters are $T$ dependent as $\Delta \approx
\Delta_0\sqrt{\tau}\,,\Delta' \approx \Delta'_0\sqrt{\tau}$ where
$\tau=(T_c-T)/T_c$; hence $\xi \approx \xi_0/\sqrt{\tau}\,,
\xi' \approx \xi'_0/\sqrt{\tau}$.

In principle the current has also a diamagnetic term $(e/c)n_{
edge}(x)A_y(x)$; the ratio of this term to the London term $(c/4\pi
\lambda^2)A_y(x)$ is $1/[(k_F\xi_0')^2\tau]$ where $\tau
=(T_c-T)/T_c$. Hence the effect of this diamagnetic current is small
except very near $T_c$, i.e. for $\tau>(k_F\xi_0')^{-2}$ (or
$\tau>(k_F\xi_0)^{-2}$ for the (110) surface). In the range
 where the order parameter fluctuations exceed its mean value mean
field breaks down; this range, which is between $\tau<1/k_F\xi_0$ in
2-dimensions and
$\tau<(1/k_F\xi_0)^4$  in 3-dimensions is excluded in our analysis.

We consider first $d+id'$; the factor $k_y\tanh (\epsilon_{\zeta}/2T)$ is
symmetric in $k_y$, therefore within the
integration in Eq. (\ref{edge}) on the $\epsilon_{\zeta} \geq 0$
range (Fig. 1) the $\zeta <0$ segment can be shifted into a $\zeta
>0$ one so that a complete $(0,\pi/2)$ range results. In terms of
the density $n=k_F^2/2\pi d$ and the $T=0$ penetration length
$\lambda_0$ ($\lambda \approx \lambda_0/\sqrt{\tau}$) where $\lambda_0^{-2}=4\pi ne^2/mc^2=2k_F^2e^2/mc^2d$
we obtain
\begin{eqnarray} \label{edges}
\frac{4\pi}{c}j_{edge}(x)&=& \frac{2\phi _0}{\pi \xi '\lambda
_0^2}\int_0^{\pi /2}d\zeta \cos \zeta \sin ^2\zeta e^{-2x\sin
\zeta /\xi '} \tanh[\frac{\Delta \cos 2\zeta +
(e/c)v_F\sin \zeta A_y(x)}{2T}]\nonumber\\
n_{edge}(x)&=& \frac{ek_F}{\pi d \xi '}\int_0^{\pi /2}d\zeta \cos
\zeta \sin \zeta e^{-2x\sin \zeta /\xi '}
\end{eqnarray}
where the rapid oscillatory $\sin ^2kx$ is replaced by its average
$\frac{1}{2}$. Note that for $\Delta =0$ or $\Delta ' =0$ all
angles $\zeta$ are allowed in the solution of Eq. (\ref{eigen0})
and then the current vanishes. This demonstrates that BTRS leads
to current carrying surface states. We note also that if the
Doppler shift $\sim A_y(x)$ is ignored the integrated current
$j_{edge}(x)$ vanishes, unlike the p wave case \cite{sigrist2}.

\begin{figure}[b]
\begin{center}
\includegraphics[scale=0.5]{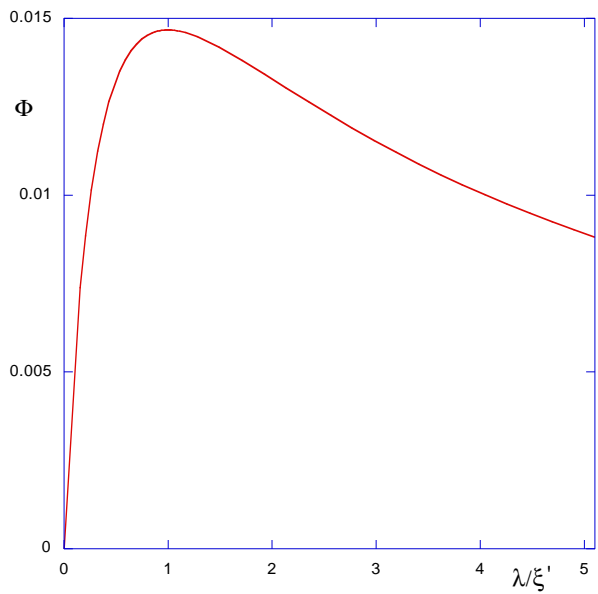}
\end{center}
\caption{Spontaneous flux for a $(100)$ boundary in thick films
 ($\xi'<{\bar d}$)
 in units of  $2\phi _0L_y\lambda \Delta/\pi \lambda
_0^2T_c$.}
\end{figure}

The response of the condensate to $j_{edge}$ involves the London
terms as well as coupling to the scalar potential at the surface;
the latter terms are small as $1/k_F\xi_0$ at low $T$ or
vanish at $T\rightarrow T_c$
(see Appendix C). London's equation with $j_{edge}(x)$ as a
source term is then,
\begin{equation}\label{london}
    -\nabla ^2 A_y(x)=[-(1/\lambda^2)A_y(x)
    +(4\pi /c)j_{edge}(x)]\theta (x) \label{London}
    \end{equation}
where $\theta (x)$ is a step function. This assumes a thick film,
i.e. no dependence on the z direction; the thin film limit is
considered below. For a thick film the condition of no external
field at $x\rightarrow -\infty$ implies $H_y(0)=0$. Eq.
(\ref{london}) is then solved by the Greens' function
\begin{equation}\label{Gbound}
G(x,x')=-(\lambda
    /2)[\exp (-|x-x'|/\lambda)+\exp (-|x+x'|/\lambda)]
    \end{equation}
which satisfies the boundary condition
$\partial_xG(x,x')|_{x=0}=0$ equivalent to
$H_y(x=0)=\partial_xA_y|_{x=0}=0$. $A_y(x)$ then satisfies an
integral equation
\begin{eqnarray}\label{Ay}
A_y(x)=\frac{\phi_0\lambda }{\pi \xi' \lambda_0^2}\int_0^{\pi
/2}d\zeta \cos \zeta \sin ^2\zeta \int_0^{\infty} &&
dx'\tanh[\frac{\Delta \cos
2\zeta+(e/c)v_F\sin \zeta A_y(x')}{2T}][e^{-|x-x'|/\lambda}+\nonumber\\
&& e^{-|x+x'|/\lambda}]e^{-2x'\sin \zeta /\xi '}
\end{eqnarray}
The doppler shift, as shown below, is significant only very near
$T_c$ or at very low temperatures. Neglecting first the Doppler
shift and at $T\rightarrow T_c$ Eq. (\ref{Ay}) becomes
\begin{equation}\label{Ay20}
A_y(0)=(2\phi _0\lambda \Delta/\pi \lambda _0^2T_c)\int
_0^{\pi/2}d\zeta \cos \zeta \sin ^2\zeta \cos 2\zeta (2\sin \zeta
+\xi '/\lambda)^{-1}\,.
\end{equation}
The total spontaneous flux is $\Phi =A_y(0)L_y$ where $L_y$ is the
length of the boundary. We consider $2\phi _0L_y\lambda \Delta/\pi
\lambda _0^2T_c$ as a flux unit, e.g. for $L_y=2cm$
\cite{polturak} and typical YBCO parameters it is $\approx 10^5
\phi_0$. This flux unit is weakly temperature dependent since
$\lambda \Delta \approx \lambda_0\Delta_0$ is finite at
$T\rightarrow T_c$. The ratio ${\tilde \Phi}=-\Phi / (2\phi
_0L_y\lambda \Delta/\pi \lambda _0^2T_c)$ is plotted in Fig. 2; it
varies between $\lambda /12\xi '$ at $\lambda \ll \xi '$ (weak
BTRS) to $\xi'/12\lambda$ at $\lambda \gg \xi '$ (strong BTRS)
with a maximum of $0.014$ at $\lambda \approx \xi '$. For a
$(110)$ surface replacing $\xi'$ by $\xi$ (considering only
$\xi\ll \lambda$) we obtain ${\tilde \Phi}=\xi\Delta '/12\lambda
\Delta$, much smaller than for a $(100)$ surface. The reason for
the dominance of the (100) surface is the steeper spectra
$\epsilon \sim \Delta$ for this case. The result that $\Phi$ is
weakly temperature dependent at $T\rightarrow T_c$ is consistent
with the spontaneous magnetization data \cite{polturak}; more
details on the data follow in section VI.

At low temperatures $T\ll T_c$ the result for the (100) surface is
of the same order as that near $T_c$ while for the (110) $\Phi$
is enhanced upon cooling, becoming at $T\ll T_c, \Delta'$ of the
order of ${\tilde \Phi}\approx \xi/\lambda$. The various limiting
forms of ${\tilde \Phi}$ are collected in table I.

\hspace{15mm}
\begin{table*}
\caption{Spontaneous flux ${\tilde \Phi}$ [flux in units of $-\phi
_0L_y\lambda \Delta/(\pi \lambda _0^2T_c)$] for various surfaces,
weak ($\xi'\gg\lambda$) or strong BTRS ($\xi'\ll\lambda$) and
various temperature limits. Comments: (i) If thin film is not
specified the entry corresponds to thick films with thickness
${\bar d}\gg \lambda, \xi'$ [(100) surface] or ${\bar d}\gg
\lambda, \xi$ [(110) surface]. (ii) All entries correspond to
$d+id'$ except the $d+is$ one which refers only to (110) surface;
for thin films it is the same as (110) $d+id'$ (except a factor -5
in the $T\rightarrow T_c$ line). (iii) $T\rightarrow T_c$ entries
for ${\tilde \Phi}$ exclude paramagnetic anomaly regions which are
given in the last column. (iv) BTRS which sets in at a temperature
$T_c'\ll T_c$ has ${\tilde \Phi}$ values corresponding to modified
temperature intervals. The only paramagnetic anomalies in this
case are (100) $T<T_s$ and (110) thin film.  (v) The fluctuation
region (e.g. $\tau\lesssim 1/k_F\xi_0$ in 2-dimensions) is excluded, hence
the region $\tau=(T_c-T)/T_c$ in the last column is relevant only if it is a larger one.}
\begin{ruledtabular}
\begin{tabular}{|c|c|c|c|c|}
 geometry & temperature & case $T_c'\ll T_c$ &
  ${\tilde \Phi}$ & paramagnetic anomaly\\ \hline
 (100) $\xi'\gg\lambda$ & $T\rightarrow T_c$ & & $\frac{\lambda}{15\xi'}$
 & $\tau \approx \left(\frac{\Delta'}{\Delta}\right)^2$ \\
  & $T\ll T_c$  & $T<T_c'$ & $(2-\sqrt{2})\frac{\lambda T_c}
  {6\xi' \Delta_0}$&  \\
  (100) $\xi'\ll\lambda$  & $T\rightarrow T_c$ & &
  $\frac{\xi'}{12\lambda}$ &
  $\tau \approx \left(\frac{\xi}{\lambda}\right)^2$ \\
   & $T\ll T_c$ & $T<T_c'$ & $(\sqrt{2}-1)\frac{\xi'T_c}
   {2\lambda \Delta_0}$ &  \\
  (110) & $T\rightarrow T_c$ & $T\rightarrow T_c'$ & $\frac{\xi^2}{12\lambda\xi'}$ &
  $\tau \approx \left(\frac{\xi}{\lambda}\right)^2$   \\
   & $T_s<T\ll T_c, \Delta'$ & $T_s<T\ll T_c'$ & $(\sqrt{2}-1)\frac{\xi T_c}{2\lambda \Delta_0}$ &  \\
  & $T<T_s$ & $T<T_s$ & $\frac{T_c}{\Delta_0}$ & $T_s\approx \frac{\xi}{\lambda}T_c$ \\
  (100) thin film & $T\rightarrow T_c$ & &  $\frac{\lambda}{15\xi'}$ &
  $\tau \approx \left(\frac{\Delta'}{\Delta}\right)^2$ \\
   & $T\ll T_c$ & $T<T_c'$ & $(2-\sqrt{2})\frac{\lambda T_c}{\xi'\Delta_0}$ & \\
  (110) thin film & $T\rightarrow T_c$ & $T\rightarrow T_c'$ &
  $-\frac{2\lambda_0 T_c}{15\xi' \Delta_0}$ & Doppler dominated \\
     & $T\ll T_c$ & $T\ll T_c'$ & $\pm \frac{2\lambda T_c}{3\xi \Delta_0}$ &
     Doppler dominated \\ $d+is$ & $T\rightarrow T_c$ & $T \rightarrow
     T_c'$ & $\frac{\xi}{2\xi_s}$ & $\tau \approx \left(\frac{\xi}
     {\lambda}\right)^2$
     \\ & $T\ll T_c, \Delta_s$ & $T\ll T_c'$ & $\frac {T_c}{\Delta_0}$ &
     $T_s \approx \frac{\xi}{\lambda}T_c$
\end{tabular}
\end{ruledtabular}
\end{table*}

We consider next the results with the Doppler shift. For the (100)
surface and $\xi'\gg\lambda$ the kernel $G(x,x')$ is localized at
$x\approx x'$ so that $A_y(x')$ can be replaced by $A_y(x)$ in Eq.
(\ref{Ay}). Near $T_c$ we expand the $\tanh $ and obtain a term
which modifies $1/\lambda^2$, i.e.
\begin{equation}
[\frac{d^2}{dx^2}-\frac{1}{\lambda^2}+\frac{\Delta'}{2T\lambda_0^2}]A_y(x)=
\frac{4\pi}{c}j_y^{(0)}(x)
\end{equation}
where $j_y^{(0)}(x)$ is the current in the absence of the Doppler
term. Very neat $T_c$, the effective London length $\lambda_{eff}$
where $1/\lambda_{eff}^2=1/\lambda^2-\Delta'/2T\lambda_0^2$
becomes imaginary so that there is no Meissner effect, i.e. a
magnetic field can penetrate into the bulk. Hence a sharp sign
change of $A_y(0)$ from paramagnetic to diamagnetic is expected at
$\tau\approx (\Delta'/\Delta)^2$. For $T\ll T_c$ we obtain
from $j_y^{(0)}$ that $\frac{e}{c}v_FA_y(0)\approx \Delta'\ll
\Delta$, i.e. the Doppler shift is negligible.

For the (100) surface and $\xi'\ll\lambda$ the $x'$ integration in Eq.
(\ref{Ay}) is limited to $\xi'$, hence we can replace $A_y(x')$ by
$A_y(0)$ to yield
\begin{equation}\label{Ay1}
A_y(x)=\frac{\phi_0\lambda}{\pi\lambda_0^2}\int_0^{\pi/2}d\zeta'
\cos \zeta \sin \zeta \tanh [\frac{\Delta \cos 2\zeta
+\frac{e}{c}v_F\sin \zeta A_y(0)}{2T}]e^{-x/\lambda} \,.
\end{equation}
At $T\rightarrow T_c$ this becomes
\begin{equation}\label{Ay2}
(1-\frac{\hbar v_F \lambda}{3T\lambda_0^2})A_y(0)=A_y^{(0)}
\end{equation}
hence the response changes sign at $\tau \lesssim
(\xi/\lambda)^2$. For $T\ll T_c$ the Doppler term can be neglected
$\frac{e}{c}v_FA_y(0)\approx \Delta_0\xi\xi'/\lambda^2\ll
\Delta_0$. For the (110) surface the form (\ref{Ay2}) applies with
$\xi'\rightarrow \xi$ (considering always $\xi\ll \lambda$) which
does not affect the left hand side of Eq. (\ref{Ay2}); hence a
paramagnetic anomaly at $\tau\approx (\xi/\lambda)^2$ .

A remarkable feature of Eq. (\ref{Ay1}) is that it allows
spontaneous magnetization for the (110) surface even if
$\Delta'=0$, as studied earlier \cite{higashitani,barash}. The
critical temperature can be deduced from Eq. (\ref{Ay1}) (with
$\Delta\rightarrow \Delta'$) by assuming a small probing $\Delta'$
and look for the $A_y$ response, which from Eq. (\ref{Ay2})
diverges at $T_s=v_F/3\lambda_0\approx (\xi/\lambda)T_c \ll T_c$.
Furthermore, at $T=0$ Eq. (\ref{Ay1}) yields
\begin{equation}\label{Ay3}
A_y(0)=\frac{\phi_0}{\pi\lambda_0}{\mbox sign} [A_y(0)]
\end{equation}
hence a spontaneous magnetization flux of $\pm
\phi_0L_y/\pi\lambda_0$.

In Fig. 3 we show the low temperature form of $A_y(0)$ for (110).
For $\Delta'=0$ it shows a spontaneous magnetization (dotted
lines) below a critical temperature, while for $\Delta'\neq 0$ it
shows enhancement near $T_s$ where it joins one of the low $T$
branches. In comparison the (100) flux depends weakly on
temperature and is much stronger than that of (110) at least at
high temperatures.
\begin{figure}
\begin{center}
\includegraphics[scale=0.5]{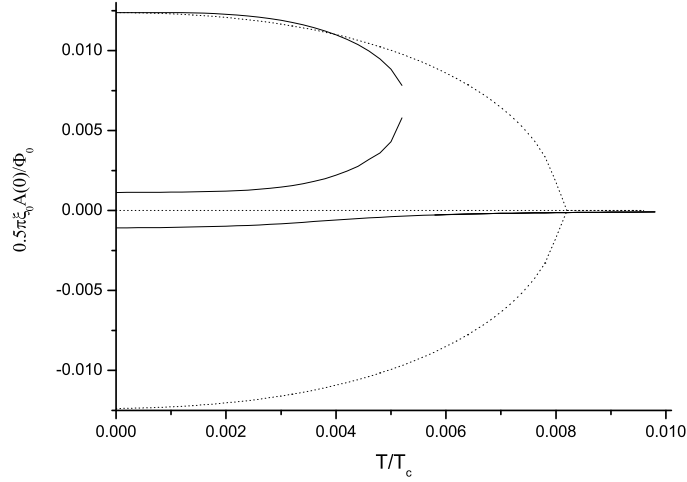}
\end{center}
\caption{ Spontaneous flux at a (110) surface for $\Delta'=0$
(dotted lines) showing a critical temperature $T_s$. For
$\Delta'/\Delta=0.01$ it shows enhancement below $T_s$.}
\end{figure}

We consider next the thin film case, for which London's equation
is
\begin{equation}\label{L2}
-\nabla ^2A(x,z)={\bar d}
[-\frac{1}{\lambda^2}A_y(x)+\frac{4\pi}{c}j(x)]\theta(x)\delta(z)
\end{equation}
where ${\bar d}$ is the film thickness.
Assuming that one can Fourier transform $A(x,z)$ into $A(q,k)$,
integration of the $k$ dependence $\sim (q^2+k^2)^{-1}$ yields for
$A_y(x)=A_y(x,z=0)$
\begin{equation}\label{L3}
A_y(x)=\int \frac{dq}{2\pi |q|}\int dx' e^{iq(x-x')} {\bar
d}[-\frac{1}{\lambda^2}A_y(x)+\frac{4\pi}{c}j(x)]\theta(x)\,.
\end{equation}
The $q$ integration then yields
\begin{equation}\label{L4}
A_y(x)-A_y(0)=\int_0^{\infty}dx' \ln \left|\frac{x-x'}{x'}\right|
{\bar d}[-\frac{1}{\lambda^2}A_y(x)+\frac{4\pi}{c}j(x)]
\end{equation}
implying a slow decay of $A_y(x)$. While a solution for $A_y(x)$
appears difficult to obtain, the value of $A_y(0)$ is readily
noticed from the boundary condition. The absence of external filed
requires a finite $H_y(x=0)$ for the thin film geometry. Hence to
avoid divergence of $dA/dx|_{x=0}$ where
\begin{equation}\label{L5}
\frac{dA(x)}{dx}|_{x=0}=\int_0^{\infty}\frac{dx'}{x'} {\bar
d}[\frac{1}{\lambda^2}A_y(x)-\frac{4\pi}{c}j(x)]
\end{equation}
one must have
$A_y(0)=\lambda^2\frac{4\pi}{c}j(0)$, i.e.
\begin{equation}\label{L7}
A_y(0)=\frac{2\lambda^2\phi_0}{\pi \xi'
\lambda_0^2}\int_0^{\pi/2}d\zeta' \cos \zeta \sin^2 \zeta \tanh
[\frac{\Delta \cos 2\zeta +\frac{e}{c}v_F\sin \zeta A_y(0)}{2T}]
\end{equation}
which interestingly has the same form as Eq. (\ref{Ay1}) except
that here it is valid for all $\xi'$. In particular, when the
Doppler shift can be neglected we obtain ${\tilde \Phi}=\lambda
/15\xi '$ which in Fig. 2 is the tangent line to the thick film
curve at the origin.
 Hence we can define two regimes: Weak BTRS
with $\lambda/\xi'<1$ where the spontaneous flux is $T$ and ${\bar
d}$
 independent, and strong BTRS with $\lambda/\xi'>1$ where film
thickness matters, with the thin film showing a stronger effect.
For strong BTRS $T$ dependence is induced as $\xi'<{\bar d}$
changes to the thin film case $\xi'>{\bar d}$ as $T\rightarrow
T_c$.

Consider now the Doppler shift for thin films; expansion near
$T_c$ yields $1/\lambda_{eff}=1/\lambda^2-\Delta'/2T\lambda_0^2$
which as above changes sign at $\tau \approx (\xi/\xi')^2$,
i.e. a paramagnetic anomaly. This temperature is the same as for
the thick film case except that here it is valid also for
$\xi'<\lambda$. Hence for $\Delta'/\Delta=0.1$ we can have an
anomaly at an accessible temperature of $(T_c-T)/T_c\approx
10^{-2}$, as shown in Fig. 4. For $T\ll T_c$ the Doppler effect is
small.
\begin{figure}
\begin{center}
\includegraphics[scale=0.5]{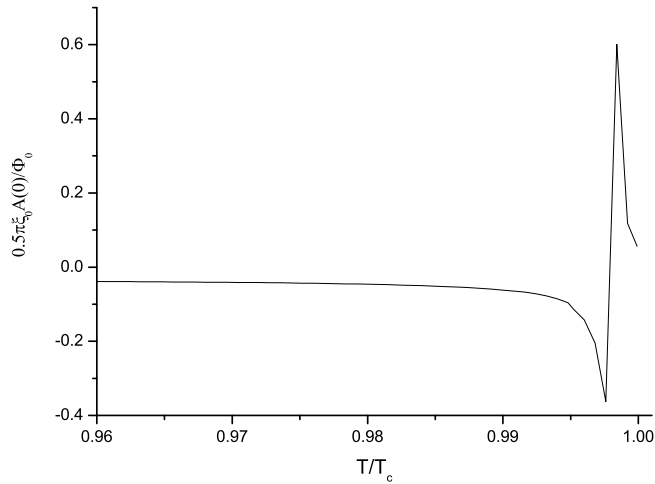}
\end{center}
\caption{Doppler induced paramagnetic anomaly near $T_c$ for thin
films with a (100) surface and $\Delta'/\Delta=0.1$ }
\end{figure}

For a (110) surface the scales of $A_y(x)$ are $\lambda, \xi$,
hence the the thin film situation applies when ${\bar d}\ll \xi$,
which is more difficult to achieve. Near $T_c$ we obtain
$A_y(0)=\frac{2\phi_0}{15\pi \xi}{\Delta'}{\Delta}>0$ which is
paramagnetic, while at $T\ll T_c$ we have $A_y(0)=\pm
\frac{4\phi_0}{3\pi \xi_0}$. A spontaneous flux even with
$\Delta'=0$ is possible also here as in the (110) thick film case.

All the various forms for the magnetization and Doppler effects
are summarized in table I. The table also considers the
possibility that BTRS sets in at a temperature $T_c'\ll T_c$. In
this case $\xi'$ diverges at $T_c'$ so that near $T_c'$  we have $\xi'\gg
\lambda$ while at $T\ll T_c'$ also $\xi'<\lambda$ is possible,
resulting in a temperature dependent ${\tilde \Phi}$ for this
strong BTRS case.

Finally we consider the $d+is$ case. Here only (110) is relevant
and $\xi\ll \lambda$ for high $T_c$ materials. The edge current is
\begin{equation}
\frac{4\pi}{c}j_{edge}(x)= \frac{2\phi _0}{\pi \xi \lambda
_0^2}\int_0^{\pi /2}d\zeta \cos \zeta \sin ^2\zeta e^{-2x\sin
\zeta /\xi } \tanh[\frac{\Delta_s + (e/c)v_F\sin \zeta
A_y(x)}{2T}]\,.
\end{equation}
Near $T_c$ Eq. (\ref{Ay20}) applies with $\Delta \cos 2\zeta$
replaced by $\Delta_s$ and $\xi'\rightarrow \xi$, hence ${\tilde
\Phi}=\Delta_s/2\Delta=\xi/2\xi_s$. At low temperatures the $\tanh
$ is replaced by $1$, leading to ${\tilde \Phi}=T_c/\Delta$. The
various results are given in table I.

\section{Quantum Hall effects - a surface approach}

In this section we study a surface formulation of quantum Hall
effects (QHE). For the usual charge conduction, in the absence of
an external field, and given that the surface fields decay in the
bulk (Meissner effect as shown in section III), Amp{\'e}re's law implies a
zero net current, i.e. a net Hall
conductance $\sigma_{xy}=0$. We focus therefore on spin and thermal
Hall effects. These were shown to be quantized first by a network
model simulations \cite{horovitz} and then by the relation to
edge states \cite{senthil}. The $d+is$ state has no surface states
near $\epsilon=0$, hence no Hall effect within linear response;
there may be a response when the voltage exceeds $\approx
2\Delta_s/e$. We consider therefore in this section only the $d+id'$
case.

The main ingredient is the chiral nature of the surface states.
These states have two branches whose spectra vanishes at $k_y
=Q\equiv \pm k_F/\sqrt{2}$. Linearizing near this point the
spectrum for $k_y=q\pm Q$ is
\begin{equation}\label{disprsion}
\epsilon =vq
\end{equation}
where $v=\sqrt{8}\Delta/k_F$ for a (100) surface while
$v=\sqrt{8}\Delta'/k_F$ for the (110) (up to a small Doppler
shift; the actual value of $v$ is not important for the eventual
result for the Hall conductance). We wish to rewrite the surface
modes in terms of two branches $\eta_{1,q},\eta_{2,q}$ where
\begin{eqnarray}\label{eta12}
\eta_{1,q}&=&\eta_{q+Q \uparrow}\nonumber\\
\eta_{2,-q}&=&\eta^{\dagger}_{q+Q\downarrow} \nonumber\\
\eta_{2,q}&=&\eta_{q-Q\uparrow}\nonumber\\
\eta_{1,-q}&=&-\eta^{\dagger}_{q-Q\downarrow}
\end{eqnarray}
with the $\pm$ signs needed for continuity (see below). Thus,
instead of two $q>0$ branches with two $\uparrow,\downarrow$ spins we
have now two branches, each of a single degree of freedom, with
both $q>0$ and $q<0$. The transformation of Eq.
(\ref{gammaspinor}) with the eigenfunctions (\ref{unorm}) is
\begin{eqnarray}\label{g}
\left( \matrix{ \Psi_{\uparrow}({\bf r}) \cr \cr
\Psi^{\dagger}_{\downarrow}({\bf r})} \right)=\sum_{q>0}\{ f_q(x)
\left(\matrix{ 1 & 1 \cr \cr -1 & 1} \right)\left( \matrix{
\eta_{1,q}e^{i(Q+q)y} \cr \cr \gamma_{2,-q}e^{-i(Q+q)y} }\right) +
  f_{-q}(x) \left(\matrix{ 1 & -1 \cr \cr 1 & 1} \right)\left(
\matrix{ \eta_{2,q}e^{-i(Q-q)y} \cr \cr \gamma_{1,-q}e^{i(Q-q)y}}
\right)\}&&\nonumber\\
&&
\end{eqnarray}

where the two terms correspond to the two branches near $\pm Q$
and
\begin{equation}\label{fq}
f_q(x)=\sqrt{\frac{2|Q+q|}{\xi'L_yk_F}}\sin
(x\sqrt{k_F^2-(Q+q)^2}\,)e^{-x|Q+q|/\xi'k_F}\,.
\end{equation}
Note e.g. that the $Q\pm q$ Fourier components of
$\Psi_{\uparrow}({\bf r})$ are $-\eta_{1,q}$ for both $\pm q$ so
that the $\pm $ signs in Eq. (\ref{eta12}) are needed for
continuity of the Fourier transform. We can therefore define
fields with continuous Fourier transforms
\begin{eqnarray}
\Psi_1({\bf r})&=&\sum_qe^{i(Q+q)y}\eta_{1,q}f_q(x)\nonumber\\
\Psi_2({\bf r})&=&\sum_qe^{i(-Q+q)y}\eta_{2,q}f_{-q}(x)\,.
\end{eqnarray}
Eq. (\ref{g}) becomes
\begin{eqnarray}
\Psi_{\uparrow}({\bf r})&=&\Psi_1({\bf r})+\Psi_2({\bf
r})\nonumber\\
 \Psi^{\dagger}_{\downarrow}({\bf r})&=&-\Psi_1({\bf
r})+\Psi_2({\bf r})
\end{eqnarray}
or in terms of spinor
\begin{eqnarray}
{\tilde \Psi}({\bf r})=\left( \matrix{ \Psi_1({\bf r}) \cr \cr
\Psi_2({\bf r})} \right)
\end{eqnarray}
we have for the spinor Eq. (\ref{gammaspinor})
\begin{eqnarray}\label{Psi2}
\Psi({\bf r})=
\left( \matrix{ \Psi_{\uparrow}({\bf r}) \cr \cr
\Psi^{\dagger}_{\downarrow}({\bf r})}
\right)=\left(\matrix{ 1 & 1 \cr \cr -1 & 1} \right){\tilde
\Psi}({\bf r})\,.
\end{eqnarray}

The kinetic energy has the form
\begin{equation}\label{HK}
{\cal H}_K=\sum_q
vq[\eta^{\dagger}_{1,q}\eta_{1,q}+\eta^{\dagger}_{2,q}\eta_{2,q}]=\int
dxdy {\tilde \Psi}^{\dagger}({\bf r})[-iv\partial_y]{\tilde
\Psi}({\bf r})\,.
\end{equation}

A crucial observation for QHE is the role of impurities  which in
general has the form
\begin{equation}\label{Himp}
{\cal H}_{imp}=\int dxdy \Psi^{\dagger}({\bf r})\tau_3
\Psi({\bf r})V({\bf r})
\end{equation}
Within the subspace of surface states this interaction becomes,
using Eq. (\ref{Psi2})
\begin{equation}\label{Himp2}
{\cal H}_{imp}=\int dxdy {\tilde \Psi}^{\dagger}({\bf r})\tau_1
{\tilde \Psi}({\bf r})V({\bf r})
\end{equation}
This impurity potential can be "gauged" away \cite{senthil} by a
transformation
\begin{equation}\label{gauge}
{\tilde \Psi}({\bf r})\rightarrow \exp [(i/v)\int^y
V(x,y')dy'\tau_1]{\tilde \Psi}({\bf r})
\end{equation}
which eliminates the impurity potential. Hence transport of chiral
states is equivalent to that of a pure system. Chirality implies no channel for
backscattering, hence impurities are indeed expected to be
ineffective. As noted in section III, (100) surface states with
energy $>\Delta'$ may mix with bulk states by impurities. The QHE
is then limited to temperatures $T<\Delta'$.

To evaluate the spin Hall conductance, we define a spin voltage
$V_s(x)$ (x is a coordinate perpendicular to the edge) such that
$(\hbar/2)\rho_s(x)V_s(x)$ is the coupling energy density to a
density $\rho_s(x)$ of $\hbar/2$ spins. This can be represented by
a Zeeman term with $V_s(x)=eB_z(x)/mc$ where $m$ is the electron
mass; the corresponding force in the $x$ direction is
$(e/mc)dB_z(x)/dx$. The unit of spin conduction, in analogy with $e^2/h$ of
the charge conduction is $(\hbar/2)^2/h=\hbar/8\pi$. Hence the
spin Hall conductance $\sigma_{xy}^{spin}=I_s/V_s$ is
\begin{equation}
\sigma_{xy}^{spin}=2\frac{\hbar}{8\pi}\mbox{sign} (\Delta \Delta
') \label{spin}
\end{equation}

The thermal Hall conduction is derived in a similar way from the
heat conduction of an ideal gas, yielding
\begin{equation}
\frac{K_{xy}}{T}=\frac{2\pi^2k_B^2}{3h} \mbox{sign} (\Delta \Delta')\label{K}
\end{equation}
Hence $K_{xy}/T$ is also quantized in this weakly disordered
system.

 We reconsider now the effect of disorder on the spin
Hall conductance. Imagine many random Hall systems, each with
their own localized chiral states which are weakly coupled. If the
couplings are too weak we expect no currents between the systems
so overall $\sigma_{xy}^{spin}=0$. As the coupling strength
increases we expect a finite current to circulate around the
ensemble of grains, leading to Eqs. (\ref{spin},\ref{K}). The
transition is in fact induced by disorder: For weak disorder the
argument of Eq. (\ref{gauge}) holds and the Hall coefficients have
their quantized values, Eqs. (\ref {spin},{\ref{K}). As
disorder increases, opposite chiral channels get coupled leading
to formation of localized chiral loops which eventually have an
insulating behavior, i.e. the Hall conduction vanishes. This
quantum Hall plateau transition \cite{horovitz} has been
simulated by a network model, showing a novel type of QHE
criticality.

\section{Quantum Hall effects - a bulk approach}

We consider next the effective action of a bulk
$d+id'$ superconductor and derive its (charge) Hall conductance
$\sigma_{xy}({\bf q},\omega)$. We assume a thin film
situation with the scalar and vector potentials $\phi,{\bf A}$ being
$z$ independent, as well as $A_z=0$. In terms of the Nambu spinors Eq.
(\ref{spinor}) the off-diagonal Hamiltonian $\int d^2r
\psi^{\dagger}({\bf r})h_{\Delta}\psi ({\bf r})$  (Eq.
\ref{hdelta}) is
\begin{equation}
 h_{\Delta }=-[\Delta
(-\partial_{x}^{2}+\partial_{y}^{2})\tau_{1}+\Delta'
\partial_{x}\partial_{y}\tau_{2}]/k_{F}^{2} \,
\end{equation}
and we neglect terms with $\nabla\theta<< k_{F}$.  The action in
presence of the electromagnetic potentials ${\bf A}, \varphi $ is then
\begin{eqnarray}
S&=&\int d^2rdt \psi^{\dagger}(i\partial_{t}-\tau_{3}
\epsilon(\hat{p})-h_{\Delta }
-\Sigma)\psi \nonumber\\
\Sigma &=&\tau_{3}(a_0+{\bf a}^2/2m) + {\bf a}\cdot {\bf p}/m- i
{\mbox{\boldmath $\nabla$}}\cdot{\bf a}/2m
\end{eqnarray}
where $\epsilon(\hat{p})=(\hat{p}^2-k_F^2)/2m$ and we introduce
the gauge invariant potentials ${\bf
a}=\frac{1}{2}{\mbox{\boldmath $\nabla$}}\theta-e{\bf A}$ and
$a_{0}= \frac{1}{2}\frac{\partial}{\partial t}\theta-e\varphi$.
The $\theta$ derivatives arise from the spinor transformation
$\Psi({\bf r})\rightarrow exp[i\tau_{3} \theta ({\bf r})/2]
\Psi({\bf r})$. Integrating out the fermions (Appendix B) and
expansion to 2nd order in ${\bf a},a_0$ leads to the effective
action

\begin{equation}\label {Seffa}
S_{eff}=\int \frac{d^{2}q d\omega} {(2\pi)^3}P_{\mu\nu}({\bf
q},\omega) a_{\mu}({\bf q},\omega ))a_{\nu}(-{\bf q},\omega))
\label{Seff}
\end{equation}
At $T=0$ and ${\bf q},\omega \rightarrow 0$ we obtain $P_{00}=N_0$
(density of states which is $N_0=m/2\pi$ in two dimensions),
$P_{ij}=-N_0c_s^2$ where $c_s=v_F/\sqrt{2}$, while $P_{0j}(q)=
i{\mbox sign}(\Delta\Delta')\epsilon_{0ij}q_{i}/(4\pi)$ and
$\epsilon_{0ij}$ is the antisymmetric unit tensor.  The latter
term reflects BTRS and is derived for $\Delta '\ll \Delta$.

Integrating out the phase $\theta$ we obtain the effective action
in terms of the electromagnetic potentials ${\bf A},\varphi$
\begin{eqnarray}
S_{eff}\{{\bf A},\varphi\}=e^2 \int \frac{d^{2}q
d\omega}{(2\pi)^3}\{&& \frac{c_{s}^2 {\bf q}^2}{c_{s}^2 {\bf
q}^2-\omega^2} [P_{00}|\varphi({\bf q},\omega)|^2- \frac{i}{4\pi}
\epsilon_{0ij}q_{i}\varphi({\bf
q},\omega)A_{j}(-{\bf q},-\omega) \nonumber\\
&&+O(\omega ^2|{\bf A}|^2)]-P_{00}(\frac{c_{s}}{c})^{2}|{\bf
A}({\bf q},\omega)|^2\}
\end {eqnarray}

The total electromagnetic action includes also the Maxwell terms
 $S_{M}=\int d^2rdt(\vec{E}^2-\vec{H}^2)/8\pi$. ${\bf A}$ may
also be integrated out, using ${\mbox{\boldmath $\nabla$}}\cdot
{\bf A}=0$ and $A_z=0$ leading to the effective action
\begin{eqnarray}
S_{eff}\{\phi)\}&=&\frac{e^2}{\chi(q,\omega)}|\phi|^2\{
[\frac{c_s^2q^2P_{00}}{c_s^2q^2-\omega^2}+\frac{q^2d}{8\pi
e^2}]\chi (q,\omega)- \frac{1}{(8\pi c)^2}\frac{c_s^4q^6}
{(c_s^2q^2-\omega^2)^2}\} \nonumber\\
\chi (q,\omega)&=&P_{00}\frac{c_s^2}{c^2}+\frac{\omega^2q^2}
{(8\pi)^2c^2P_{00}(c_s^2q^2-\omega^2)}+\frac{d\omega^2} {8\pi
c^2e^2}-\frac{dq^2}{8\pi e^2}
\end{eqnarray}
The coefficient of $|\phi|^2$ vanishes when $q\rightarrow 0$ at the
plasma frequency $\omega_p=(4\pi ne^2/m)^{1/2}=c/\lambda_0$; there
are no acoustic plasmons.

 The Hall
current $J_y$ is identified by a functional derivative with
respect to $A_y$ leading to the Hall coefficient
\begin{equation}
\sigma_{xy}({\bf q},\omega)={\mbox
sign}(\Delta\Delta')\frac{e^2}{4\pi \hbar}
\frac{c_{s}^{2}q^2}{c_{s}^{2}q^{2}-\omega^{2}} \label{xy}
\end{equation}
Transport is defined by taking the $q\rightarrow 0$ limit first,
i.e. $\sigma_{xy}=0$.  Hence the conventional Hall coefficient
vanishes, as expected from Galilean invariance \cite{read}.  A
limit in which $\omega \rightarrow 0$ is taken first yields a
quantized "static" conductance $e^2/2h$ which was argued to
correspond to $\sigma_{xy}\neq 0$ in presence of a boundary
\cite{goryo}.  In absence of an external magnetic field, and given
a spontaneous magnetization decaying in the bulk (as shown in section
III), Amp{\'e}re's law yields zero total current, hence
$\sigma_{xy}=0$; this is valid also with a boundary and an external
electric field. It is intriguing, however, that $\sigma_{xy}({\bf
q},\omega)$ has a nontrivial structure and space resolved
measurement of a Hall current could then probe the full Eq.
(\ref{xy}). We note that a result similar to Eq. (\ref{xy}) was
obtained for superfluid $^3$He \cite{goryo2}.

\section{Conclusions}

We consider now in more detail the experimental data on the
spontaneous magnetization \cite{polturak}.  The data shows that for
a YBCO disc with a perimeter of $L_y\approx 2$cm the spontaneous
magnetization is temperature independent in the range 80-89K and is
also thickness independent in the range 30-300nm with a value of
$\approx$37$\phi_0$.  Taking $\lambda \Delta \approx
\lambda _0\Delta _0$, their $T=0$ value, and typical YBCO
parameters we find ${\tilde \Phi }\approx 10^{-3}$.  The temperature
and thickness independence indicate weak BTRS with
$\xi '>\lambda$. For
either thick or thin films we estimate
 $\lambda/\xi '\approx 10^{-2}$ or
$\Delta'/\Delta\approx 10^{-4}$.  We propose therefore that
increasing the ratio $\Delta'/\Delta$, e.g. by using overdoped
YBCO \cite{deutscher}, one can enhance the spontaneous
magnetization up to a maximum of $\approx 10^3\phi_0$ when
$\Delta'/\Delta \approx 0.01$ within the thick
film regime.

For strong BTRS, $\lambda/\xi'>1$, the film thickness
matters, i.e. we expect a temperature
dependence due to the crossover from thick to thin film regimes at
${\bar d}\approx\xi'$ as
$T\rightarrow T_c$.  For thin films (${\bar d}<\xi'<\lambda)$ we
obtain ${\tilde \Phi}=\lambda /12\xi'$, i.e. for
YBCO the total flux can reach $10^5\Delta'/\Delta\phi_0$ per cm of
boundary, much higher than thick film values. The situation of a
strong BTRS with thin films is interesting also as being the most
likely one to show the
paramagnetic anomaly (Fig. 4) at a temperature $\approx T_c[1-(\xi/\xi')^2]$.

In conclusion, we have shown that surface states of a
$d+id'$ superconductor lead to spontaneous
magnetization which is $T$ independent and thickness independent
for weak BTRS, $\lambda/\xi'<1$, in accord with the
data \cite{polturak}.  For strong BTRS with $\lambda/\xi'>1$,
 as expected in overdoped YBCO \cite{deutscher} , a
crossover from thick to thin film behavior can lead to $T$ and
thickness dependence, as well as to an observable paramagnetic
anomaly near $T_c$.  We have shown gapless chiral surface states
for the $d+id'$ state which lead to quantization of the spin and
thermal Hall conductances. The $d+is$ state has surface currents
only at the (110) surface; its surface excitations have a gap and
therefore insulating; i.e. no nontrivial quantization of Hall
conductances.  For the charge  Hall conductance we find a
vanishing transport value, however the structure of
$\sigma_{xy}({\bf q},\omega)$ has an unusual form which exhibits
the Goldstone mode of the superconductor.

\begin{acknowledgments}
We thank Y. Dagan, G. Deutscher, A. J. Legget, O. Milo, and E.
Polturak for valuable discussions. This research was supported by
THE ISRAEL SCIENCE FOUNDATION founded by the Israel Academy of
Sciences and Humanities.
\end{acknowledgments}

\vspace{3cm}
\appendix

\section{Hamiltonian for d wave superconductor}

We derive here the interaction term for a d+id' superconductor. A
general interaction Hamiltonian in terms of a Nambu spinor Eq.
(\ref{spinor}) is
\begin{equation}\label{Hint}
{\cal H}_{int}=-\int \Psi^{\dagger}({\bf r}_1)\tau_3\Psi({\bf
r}_1)\Psi^{\dagger}({\bf r}_2)\tau_3\Psi({\bf r}_2)V({\bf
r}_1-{\bf r}_2)
\end{equation}
The order parameter has the form
\begin{equation}\label{Delta}
e^{i\theta ({\bf r})\tau_3}\tau_1 \Delta({\bf r}_1,{\bf
r}_2)=\langle \Psi({\bf r}_1)\Psi^{\dagger}({\bf r}_2)\rangle
V({\bf r}_1-{\bf r}_2)
\end{equation}
where the phase $\theta({\bf r})$ depends only on the center of
mass coordinate ${\bf r}=({\bf r}_1+{\bf r}_2)/2$. The factor
$\Delta({\bf r}_1,{\bf r}_2)$ may be complex, however, its real
and imaginary components are determined by the interactions and
their ratio is not allowed to vary in space. A d wave
superconductor is defined by a momentum dependence $k_x^2-k_y^2$
for the relative coordinate, i.e.
\begin{equation}\label{Dxy}
\Delta({\bf r}_1,{\bf r}_2)=\Delta({\bf r})\int e^{i{\bf k}\cdot
({\bf r}_1-{\bf r}_2)}(k_x^2-k_y^2)\frac{d^2k}{(2\pi k_F
)^2}=\Delta({\bf
r})(-\partial^2_{\xi}+\partial^2_{\eta})\delta^2({\mbox{\boldmath
$\rho$}})/k_F^2
\end{equation}
where the relative coordinate is ${\mbox{\boldmath $\rho$}}={\bf
r}_1-{\bf r}_2=(\xi,\eta)$. The mean field Hamiltonian is then
\begin{equation}
{\cal H}_{int}^{MF}=\int d^2rd^2\rho \Psi^{\dagger}({\bf
r}+\frac{1}{2}{\mbox{\boldmath $\rho$}})e^{i\theta({\bf
r})\tau_3}\tau_1 \Psi({\bf r}-\frac{1}{2}{\mbox{\boldmath
$\rho$}})\Delta({\bf
r})[\partial^2_{\xi}-\partial^2_{\eta}]\delta^2({\bf r})/k_F^2
\end{equation}
After partial integrations,
\begin{eqnarray}
{\cal H}_{int}^{MF}=&&\frac{1}{4}\int d^2r \Psi^{\dagger}({\bf
r})[(\partial^2_x-\partial^2_y)\Delta({\bf
r})e^{i\theta({\bf r})\tau_3}]\tau_1 \Psi({\bf r})/k_F^2\nonumber\\
&& -\int d^2r \Psi^{\dagger}({\bf r})[\partial_x\Delta({\bf
r})e^{i\theta({\bf r})\tau_3}\partial_x -
\partial_y\Delta({\bf r})e^{i\theta({\bf
r})\tau_3}\partial_y]\tau_1\Psi({\bf r})/k_F^2
\end{eqnarray}
where in the first term $(\partial^2_x-\partial^2_y)$ operates
only within the $[]$ brackets. A d' component corresponds to
$\Delta'({\bf r})
\partial_{\xi}\partial_{\eta}\delta^2({\mbox{\boldmath $\rho$}})$
and similar analysis can be followed. We assume here that all
gradients are small, i.e. $|{\mbox{\boldmath $\nabla$}}\theta|,
|{\mbox{\boldmath $\nabla$}}\Delta|/\Delta \ll k_F$, hence with
the transformation $\Psi({\bf r})\rightarrow exp[i\tau_{3} \theta
({\bf r})/2] \Psi({\bf r})$ yields the off-diagonal Hamiltonian
$\int d^2r \Psi^{\dagger}({\bf r})h_{\Delta}\Psi ({\bf r})$ where
\begin{equation}\label{hdelta}
 h_{\Delta }=-[\Delta
(-\partial_{x}^{2}+\partial_{y}^{2})\tau_{1}+\Delta'
\partial_{x}\partial_{y}\tau_{2}]/k_{F}^{2} \,
\end{equation}

The issue of gauge invariance is of some interest. The full
interaction form Eq. (\ref{Hint}) is manifestly invariant under
$\Psi({\bf r})\rightarrow exp[i\tau_{3} \int^{{\bf r}}A ({\bf
r}')\cdot d{\bf r}'] \Psi({\bf r})$. Wether the mean field form is
also gauge invariant is a matter of some debate
\cite{lee,tesanovich}. From the definition Eq. (\ref{Dxy}) it
seems that
\begin{equation}
\Delta({\bf r}_1,{\bf r}_2)\rightarrow exp[i\tau_{3} \int^{{\bf
r}_1}A ({\bf r}')\cdot d{\bf r}'] exp[i\tau_{3} \int^{{\bf r}_2}A
({\bf r}')\cdot d{\bf r}']\Delta({\bf r}_1,{\bf r}_2)
\end{equation}
and then ${\cal H}_{int}^{MF}$ is gauge invariant without having
explicit $A({\bf r})$ dependent terms. This, however, implies that
the $\partial_x,\partial_y$ terms do not follow the usual
substitution law as in the kinetic term. For the present work this
issue is irrelevant since we neglect these terms altogether, i.e.
we assume $|{\mbox{\boldmath $\nabla$}}\theta|, |{\mbox{\boldmath
$\nabla$}}\Delta| \ll k_F$.

\section{Derivation of $P_{ij}$}
We derive here an effective action for a $d+id'$ superconductor in
terms of the gauge invariant potentials $a_{\mu}(q,\omega)$, Eq. (\ref
{Seffa}).
 Integrating out the fermionic
variables in the partition function we arrive to the following action:
\begin{eqnarray}
Z&=&\int D\Phi e^{iS}\\
S(\Phi)&=&-iTr\ln\hat{G}^{-1}, \,\,\,
G^{-1}=G^{-1}_{0}-\Sigma\nonumber\\
G^{-1}_{0}&=&i\partial_{t}-\tau_{3} \epsilon(p)-h_{\Delta }
\nonumber
\end{eqnarray}

 We are interested in the long wavelength limit; also
 the  order parameter is taken at the extremum of the
effective action with only phase
 fluctuations. We retain the
first and the second order in $\Sigma$ to derive an expansion of the effective
action in the fluctuating fields $a_{\mu}(q,\omega)$.
The expansion corresponds to a one loop calculation with the
coefficients $P_{\mu,\nu}$ in Eq. (\ref{Seffa}) given by (latin
indices stand for space coordinates)
\begin{eqnarray}
P_{00}({\bf q}\omega )&=&\frac{i}{2} T\sum_{{\bf p},\omega'}
Tr[G(p,\omega')\tau_{3}G(p+q,\omega'+\omega )\tau_{3}]\\
P_{ij}{\bf q}\omega )&=&\frac{i}{2m^{2}}T \sum_{{\bf p},\omega'}
Tr[G(p,\omega')G(p+q,\omega'+\omega)p_{i}(p+q/2)_{j}]-\frac{n}{2m}\delta_{i,j}\\
P_{0j}({\bf q}\omega )&=&\frac{i}{2m}T\sum_{{\bf p},
\omega'}Tr[G(p,\omega')\tau_{3}G(p+q,\omega'+\omega)p_{j}]
\end{eqnarray}

 The diagonal time polarization operator $P_{00}$ depends
 weakly on
temperature and therefore in the limit  of small momentum and
frequency $q\rightarrow 0,\omega \rightarrow 0$ is given by
its $T=0$ value, i.e. the mean-field compressibility, $P_{0
0}(q)=N_{0}$ . The space components $P_{i,j}$ include
the diamagnetic term and paramagnetic current correlator. In the
limit $q\rightarrow 0,\omega \rightarrow 0$ they give the
mean-field superfluid stiffness; at $T\rightarrow 0$
$P_{ij}(q)=-N_{0}c_{s}^2 $ where $c_{s}=v_{F}/\sqrt{2}$.
Of special significance is the off-diagonal polarization bubble
 $P_{0,j}$ which is responsible for the Hall effect. It is
 a topological effect depending
(at least at small values of the $d_{x,y}$ order parameter) only on
the sign
of $\Delta'$.  In the same long wave-length limit we have
$P_{0j}(q)=isign(\Delta\Delta')\epsilon_{0ij}q_{i}/(4\pi)$
 where $\epsilon_{0ij}$ is
the antisymmetric unit tensor.

\section{Effective action with boundary}
We study here the Hall term with boundary and show that its effect on
London's equation is small at either $T=0$ or $T\rightarrow T_c$. The
electromagnetic response to the surface charge and currents couples
in general the vector and scalar potentials $ {\bf A}, \varphi$ with
the Hall coefficient. We estimate this effect first at $T=0$. The Hall
term relates the current along the surface ($y$ direction) and the
electric field $\partial \varphi /\partial x$ in the $x$ direction, i.e.
\begin {equation}\label {Aphis}
(\frac{1}{\lambda_{L}^{2}}-\frac{\partial^{2}}{\partial
x^{2}})A_{y}(x)-\frac{4\pi}{dc}\sigma_{xy}\frac{\partial
\varphi}{\partial x}=-\frac{4\pi }{c}j_{edge}(x)
\end{equation}
The equation for $\varphi(x)$ involves the Debye screening length
$\lambda_d=1/\sqrt {4me^2}\ll \xi',\xi$,
\begin {equation}
(\frac{1}{\lambda_d^{2}}-\frac{\partial^{2}}{\partial
x^{2}})\varphi(x)=4\pi n_{edge}(x)
\end{equation}
The Hall term is neglected here as we wish to estimate the lowest
order effect. The
solution with $\partial \varphi /\partial x=0$ involves the Greens'
function Eq. (\ref {Gbound}); at $x\ll \xi'$ it has $\partial \varphi /\partial
x\sim \lambda_d/\xi'$ while at $x\gg \lambda_d$
\begin{equation}
 \varphi(x)= -\frac {\lambda_d^2ek_F}{2\pi d \xi'}
 \int_0^{\pi /2}d\zeta \cos \zeta \sin ^2\zeta e^{-2x\sin
\zeta /\xi '}
 \end{equation}
 The ratio of $\partial \varphi /\partial x$ and $j_{edge}$ terms in
 Eq. (\ref {Aphis}) is then $1/(32\pi k_F\xi')\ll 1$; for a (110)
 surface replace $\xi'\rightarrow \xi$. Since $k_F\xi'>k_F\xi\gg 1$ is
 the criterion  for excluding the order parameter fluctuations very
 near $T_c$, we can neglect the Hall term in London's equation (\ref
 {Aphis}).

 We consider next $T\rightarrow T_c$.  The
polarization function $P_{00}$ (B2)  is obtained by replacing
$N_{0}\rightarrow N(T)$ where
\begin{equation}
  N(T)=\sum_{p}\frac{|\Delta|^{2}}{2E^{3}}tanh(\frac{E}{2T})
\end{equation}
 with $E=\sqrt{\epsilon^{2}(p)+g_1^2 + g_2^2}$ where
 \begin{eqnarray}
 g_1&=&\Delta ({\bf r})\cos 2 \zeta \nonumber\\
 g_1&=&\Delta'({\bf r})\sin 2 \zeta  \,.
  \end{eqnarray}
 The polarization function (B3) defines the temperature dependent
 London penetration depth
 $1/\lambda^{2}_{L}(T)$.

We consider in more detail the Chern Simon (or Hall) part of the action
which is the product of scalar and vector potentials
(Eq. B4).  We consider
 a  superconductor that occupies the half space $(x>0)$ where
 the order parameters $\Delta, \Delta'$ may become functions of
 $x$. The Chern Simon part of the
action can be written in configuration space in the form
\begin{eqnarray}
S_{c-s}&=&e^2 T\sum_{\omega}\int dr[b_{1}(rr')a_{0}(r\omega
)a_{y}(r\omega )+ b_{2}(rr')a_{0}(r\omega)\frac{\partial
a_{y}(r\omega )}{\partial
x}]_{r\rightarrow r'}\\
b_{1}(rr')&=& \frac{\epsilon_{ij}}{2m}\sum_{\bf
p}p_{y}\frac{\partial }{\partial
x}(F_{p}(r,r')g_{i}(r))\frac{\partial g_{j}·}{\partial
p_{x}}\\
b_{2}(rr')&=& \frac{\epsilon_{ij}}{2m}\sum_{\bf p}p_{y}
F_{p}(r',r)g_{i}(r') \frac {\partial g_{j}(r)}{\partial p_{x}}\\
F_{p}(rr')&=&\frac{2[E'th(\frac{E}{2T})-E
th(\frac{E'}{2T})]}{EE'(E'^{2}-E^{2})}
\end {eqnarray}
where $E=E(r),E'=E(r')$. Similarly as for the infinite system we
can integrate out the Goldstone mode
\begin{eqnarray}
 S_b\{{\bf A},\varphi\} & = & e^2 \int dr[P_{00}\varphi^2({\bf r})-
\frac{c}{4\pi}( \frac{1}{\lambda_{L}(T)})^{2}{\bf A}^2({\bf r})
+b_1({\bf r}) \varphi({\bf r}) A_{y}({\bf r}) +b_2({\bf r})
\varphi({\bf r}) \frac{ \partial A_{y}({\bf r})}{\partial x}]
\end{eqnarray}
In this equation we took the polarization function at zero
frequency which is legitimate for finite system (the effective
momentum deviates from zero). The coefficients $b_{1}=b_{1}(rr')|_{r\rightarrow r'}$,
$b_{2}=b_{2}(rr')|_{r\rightarrow r'}$ are given as
\begin{eqnarray}
4\pi e^{2}b_{1}&=&\frac{e^2}{\hbar cd\xi}f_{1}(x,T) \\
4\pi e^{2}b_{2}&=&\frac{e^2}{\hbar cd}f_{2}(x,T) \,.
\end{eqnarray}
 The function
$f_{1}$ appears only for a system with boundary and depends on the
space derivative of the order parameters, while $f_{2}$ at $T=0$
is the same as for an infinite superconductor . We calculate these
function for $T=0$ and for $T\rightarrow T_{c}$,
\begin{eqnarray}
f_{1}(x,T)=&\xi \frac{d}{dx}ln[\Delta
(\Delta')^{(1+\delta )/2}]  \qquad & if\,\,T=0 \\
  =&0.11\xi\frac{\Delta \Delta'}{T_{c}^{2}}\frac{d}{dx}ln[\Delta
\Delta']  & if\,\,T\rightarrow T_{c} \\
f_{2}(x,T)=&1  & if\,\,T=0 \nonumber\\
 =&0.21\frac{\Delta \Delta'}{T_{c}^{2}}  &  if\,
T\rightarrow T_{c}
\end{eqnarray}
with $\delta=2/(1+\Delta'/\Delta)$ and $\Delta(x),
\Delta'(x)>0$ is assumed.
Thus, in the limit $T\rightarrow T_{c}$ we we can write
$\lambda\approx \lambda_0/\sqrt{\tau}$, $b_1({\bf r})=
0.11(\Delta \Delta'/T_{c}^{2})\frac{d}{dx}ln[\Delta \Delta']/2hcd$
and $b_2({\bf r})= 0.21(\Delta \Delta'/T_{c}^{2})/2hcd$.

The next step involved  the equations for electromagnetic
potentials - generalized London equations by the variations of the
total action (including the Maxwell part) over these potentials.
We consider here the half-plane geometry, i.e. the superconductor
occupies the $x>0$ half-plane.
 The nonzero electromagnetic potentials $A_y, \varphi$ obey the
equations
\begin{eqnarray}
(\frac{1}{\lambda_{d}^{2}}-\frac{\partial^{2}}{\partial
x^{2}})\varphi-\frac{4\pi e^2}{c\hbar}b_{2} \frac{\partial
A_{y}}{\partial x}-\frac{4\pi e^2}{c\hbar}b_{1}
A_{y}&=& 4\pi en_{edge}(x) \\
(\frac{1}{\lambda_{L}^{2}}-\frac{\partial^{2}}{\partial
x^{2}})A_{y}-\frac{4\pi e^2}{c\hbar}b_{2}\frac{\partial
\varphi}{\partial x}+\frac{4\pi e^2}{c\hbar}(b_{1}-\partial_{x}
b_{2}) \varphi &=&-\frac{4\pi }{c}j_{edge}(x)    \label{Londonfull}
\end{eqnarray}
where $n_{edge}, j_{edge}$ are the edge charge and edge current
densities (Eq.  \ref {edges} ).

Using the expressions above for $b_1,b_2$ we find that the Chern-Simon term affects the
spontaneous magnetization, leading to an additional flux $\sim
(\Delta '/\Delta) \tau$ which vanishes at $T\rightarrow T_c$, i.e.
it is negligible compared with the other terms in Eq. (\ref{Londonfull})
which lead to constant magnetization as $T\rightarrow T_c$, as shown
in section III.

\end{document}